\documentclass[sigconf,nonacm, natbib=false]{acmart}
\thanks{© {Evgeny Dedov} {2025}. This is the author's version of the work. It is posted here for your personal use. Not for redistribution. The definitive Version of Record was published in Proceedings of the 2025 International ACM SIGIR Conference on Innovative Concepts and Theories in Information Retrieval (ICTIR) (ICTIR ’25): \url{https://doi.org/10.1145/3731120.3744587} }

\RequirePackage[
  datamodel=acmdatamodel,
  style=acmnumeric,
  sorting=none
  ]{biblatex}
\usepackage{hyperref}
\usepackage{cleveref}
\usepackage{adjustbox}
\usepackage{appendix}

\crefname{figure}{Fig.}{Figs.}
\crefname{table}{Tab.}{Tabs.}

\title{JointRank: Rank Large Set with Single Pass}
\author{Evgeny Dedov}
\affiliation{
  \institution{JetBrains}
  \city{Limassol}
  \country{Cyprus}
}
\email{evgeny.dedov@jetbrains.com}
\orcid{0009-0000-2843-6023}

\begin{document}

\begin{abstract}
Efficiently ranking relevant items from large candidate pools is a cornerstone of modern information retrieval systems -- such as web search, recommendation, and retrieval-augmented generation. Listwise rerankers, which improve relevance by jointly considering multiple candidates, are often limited in practice: either by model input size constraints, or by degraded quality when processing large sets. We propose a model-agnostic method for fast reranking large sets that exceed a model input limits. The method first partitions candidate items into overlapping blocks, each of which is ranked independently in parallel. Implicit pairwise comparisons are then derived from these local rankings. Finally, these comparisons are aggregated to construct a global ranking using algorithms such as Winrate or PageRank.

Experiments on TREC DL-2019 show that our method achieves an nDCG@10 of 70.88 compared to the 57.68 for full-context listwise approach using gpt-4.1-mini as long-context model, while reducing latency from 21 to 8 seconds.

The implementation of the algorithm and the experiments is available in the repository: https://github.com/V3RGANz/jointrank
\end{abstract}



\maketitle

\section{Introduction}

Ranking relevant items from a large candidate pool is a cornerstone task in information retrieval, powering applications ranging from web search and recommendation systems \cite{recsyssurvey, recsysgen} to retrieval-augmented generation (RAG) \cite{ragsurvey}. While powerful reranking models, particularly listwise approaches, can significantly improve relevance by considering the context of multiple candidates simultaneously, they face significant challenges when dealing with large candidate sets. Processing hundreds or even thousands of items often exceeds the input context limits of these models or, more critically, introduces prohibitive latency due to the computational cost per item or the need for multiple sequential model inferences. Although modern LLMs support large input size limits, their ability to effectively process lengthy contexts remains limited for various reasons \cite{liu2023lostmiddlelanguagemodels, yu2024defenserageralongcontext, jacob2024drowningdocumentsconsequencesscaling} \cref{tab:trec2019top1000}.

Existing strategies to handle large sets often involve iterative processes. Methods like sliding-window ranking \cite{rankgpt}, setwise heapsort  \cite{setwise}, tournament-based selection \cite{tournament_rank}, or dynamic partitioning \cite{top_down} reduce the computational burden compared to naive sorting, but still fundamentally rely on multiple sequential calls to the reranking model.

In real-time applications where minimizing the time-to-first-token or providing an immediate response is paramount, the cumulative latency from these sequential calls remains a significant bottleneck, often rendering such approaches impractical.

This paper introduces \textbf{JointRank}, a novel, model-agnostic method designed explicitly to rerank large candidate sets with high accuracy while minimizing latency. The core idea is to perform ranking in a single parallel pass. JointRank partitions the initial set of candidates into multiple, carefully constructed, overlapping blocks. Each block -- small enough to be effectively processed -- is handled independently and in parallel by the chosen listwise ranker. 

Our primary contribution is an efficient reranking approach that decouples the number of candidates from the number of sequential model calls. JointRank achieves end-to-end latency dominated by just a single parallel batch of reranker inferences, irrespective of the initial sequence length (assuming sufficient parallel processing capability). We demonstrate through experiments on synthetic data and the TREC DL-2019 \cite{trec2019} benchmark that JointRank significantly reduces latency compared to state-of-the-art iterative methods, while maintaining competitive ranking effectiveness, particularly showcasing robustness when handling large, potentially unordered candidate sets where simpler full-context approaches falter.

\section{The Ranking problem}

Ranking is a permutation of a set $\mathcal{C} = \{ D_1, D2, \ldots, D_n\}$ with respect to the query $q$. In the context of retrieval, given documents and user query, a ranking model outputs a top-$k$ subset of documents sorted by their relevance to $q$.

\paragraph{Ranker Models}

Ranking models for information retrieval are typically categorized as follows:
\begin{itemize}
    \item \textbf{Pointwise models:} Independently assign a relevance score to each (query, document) pair.
    \item \textbf{Pairwise models:} Directly compare the relevance of two documents given a query, i.e., (query, document$_1$, document$_2$).
    \item \textbf{Listwise models:} Simultaneously rank a set of documents, i.e., (query, document$_1$, document$_2$, …, document$_n$).
    \item \textbf{Setwise models:} \cite{setwise} Select the single most relevant document from a set of $N$ candidates, i.e., (query, document$_1$, document$_2$, …, document$_n$).
\end{itemize}

\textbf{Pointwise} models are commonly used in first-stage retrieval due to parallelizability (the evaluation of each document is independent). However, these models often suffer from ranking bias since each document is scored independently without context from other documents.

Ranking contexts like those used in \textbf{pairwise}, \textbf{listwise}, and recently introduced \textbf{setwise} models can yield more accurate rankings. Among these approaches, listwise models are typically the most accurate due to their holistic view of the candidate set \cite{setwise}. They require more computational power for this improvement become noticeable and generally have limited scalability with a restricted input size. The accuracy and efficiency of listwise models degrade when processing large candidate sets \cite{jacob2024drowningdocumentsconsequencesscaling} \cref{tab:trec2019top1000}.

A ranking of the full set is referred to as \textit{complete}, while a subset ranking is referred to as \textit{partial} or \textit{incomplete}. Producing a ranking from several rankings is termed \textbf{rank aggregation} \cite{rankagg, borda}, and the resulting aggregated ranking is often called the \textit{consensus} ranking.

\section{Related Work}

\subsection{LLM Ranking}

LLMs currently achieve state-of-the-art performance across various tasks, including information retrieval \cite{liang2023holisticevaluationlanguagemodels, rankgpt}. However, a notable drawback is their significant computational and time requirements. Several works have explored various zero-shot ranking approaches: \textit{Pointwise}, including binary \cite{liang2023holisticevaluationlanguagemodels} and fine-grained scoring methods \cite{Zhuang2023BeyondYA}; \textit{Pairwise} prompting, as investigated in \cite{llmpairwise}; and \textit{Listwise} ranking techniques proposed in \cite{rankgpt, lrl}. Additionally, fine-tuning approaches have been examined in studies such as \cite{rankzephyr, rankvicuna, first, lit5rank}.

\subsection{Handling large sets of documents}

\paragraph{Calibration} 

\cite{ren2024selfcalibratedlistwisererankinglarge} proposed a fine-tuning approach by training a listwise model to output calibrated scores that are comparable across different outputs. This enables splitting the initial sequence into batches, processing the batches independently, and then aggregating the rankings using these relevance scores. This is beyond our scope, as we are investigating a model-agnostic approach suitable for zero-shot reranking. However, it could potentially be used in combination with our method.

\paragraph{Sorting}

Traditional sorting algorithms (like quicksort or mergesort), which utilize pairwise comparisons, are effective theoretically but do not suit us due to the high number of sequential calls to the ranking model \cite{llmpairwise} ($N \cdot \log_2 N$). Sorting can, however, be heavily parallelized. A radical example is ranking all possible pairs simultaneously, as in PRP-AllPair \cite{llmpairwise}. If we limit the number of parallel calls (\(p\), representing the number of parallel processing units), running \(N(N-1)/2\) parallel rankings for large candidate sets may become infeasible due to throttling or insufficient computational resources. In such cases, the span time complexity of ranking all pairs reduces to \(O\left(\frac{N^2 - N}{2p}\right)\). There are also implementations of sorting algorithms designed for parallel operations, such as parallel merge sort, which achieves a span complexity of \(O(\log^3 n)\) when \(p = N\) \cite{leiserson1994introduction}. An additional factor reducing processing time is the number of candidates handled by the listwise/setwise ranker.
Moreover, ranking is not purely a sorting problem, as the focus is typically on identifying the top-\(k\) items rather than sorting the entire set.
\cite{setwise} proposes an optimized sorting approach that leverages listwise or setwise rerankers combined with dynamic pruning, which avoids processing elements that are clearly not candidates for the top-$k$ set. However, this method still entails repeated reranker inferences. In this research, we compare only with the Setwise.heapsort proposed by \cite{setwise} and use their source code to reproduce the result.

\paragraph{Tournament Approaches}
\cite{tournament_rank} propose a tournament-style approach:
\begin{enumerate}
    \item \textbf{Divide:} The $N_k$ elements are divided into groups.
    \item \textbf{Rank:} The elements within each group are ranked.
    \item \textbf{Select:} The top-$m_k$ elements from each group are selected to move on to the next stage.
    \item \textbf{Repeat:} Steps 1--3 are repeated.
\end{enumerate}

$r$ tournaments are run in parallel and their scores are aggregated.
The number of sequential calls and the span time depend on the number of selection stages $K-1$.

\paragraph{Sliding-window listwise ranking}  
Used by \cite{rankgpt}, a fixed-size ($w$) window moves from the bottom of the list to the top, iteratively promoting highly relevant documents. With step size $s$, ranking $n$ documents requires about $O(n/s)$ sequential calls, ensuring that strongly relevant documents rise rapidly upwards.

\paragraph{Top-down partitioning}  

Introduced in \cite{top_down}, TDPart ranks the $N$ candidates with a base model (BM25 \cite{bm25} or cosine similarity), and then reranks the first $w$ elements from this selection. Subsequently, the $k$'th element is chosen as the pivot element.

Next, the algorithm splits the other candidates from $k+1$ to $N$ into batches, adding the pivot element to each batch. The batches are then reranked in parallel. Any elements ranked lower than the pivot are considered irrelevant. All the other elements have a higher relevance than the pivot, so they should be compared to the first $k$ elements. Finally, the algorithm finds the new pivot element and repeats the process.

This approach also incorporates dynamic pruning, where elements that cannot possibly appear in the final top-k list are discarded, thereby reducing the computational overhead.
While more efficient, these iterative methods also require sequential reranker calls, but they often become feasible due to increased parallelization.

\section{Proposed Algorithm}

\subsection{Motivation}

In real-time applications -- for example, retrieval-augmented generation (RAG) in LLM assistants, code completion tools, or interactive Q\&A systems -- latency is a critical factor. The time-to-first-token generation must be minimized to ensure optimal responsiveness and user satisfaction. Even an additional second of latency can become a barrier that prevents the practical adoption of a given method.
All LLM-based reranking approaches that involve multiple sequential model calls become increasingly impractical as the number of calls grows, due to the cumulative latency.
Thus, there is a clear need for an efficient reranking approach that maintains high accuracy without significantly increasing latency.

\subsection{Algorithm Description}

\begin{figure}[h!]
    \centering
    \includegraphics[width=\linewidth]{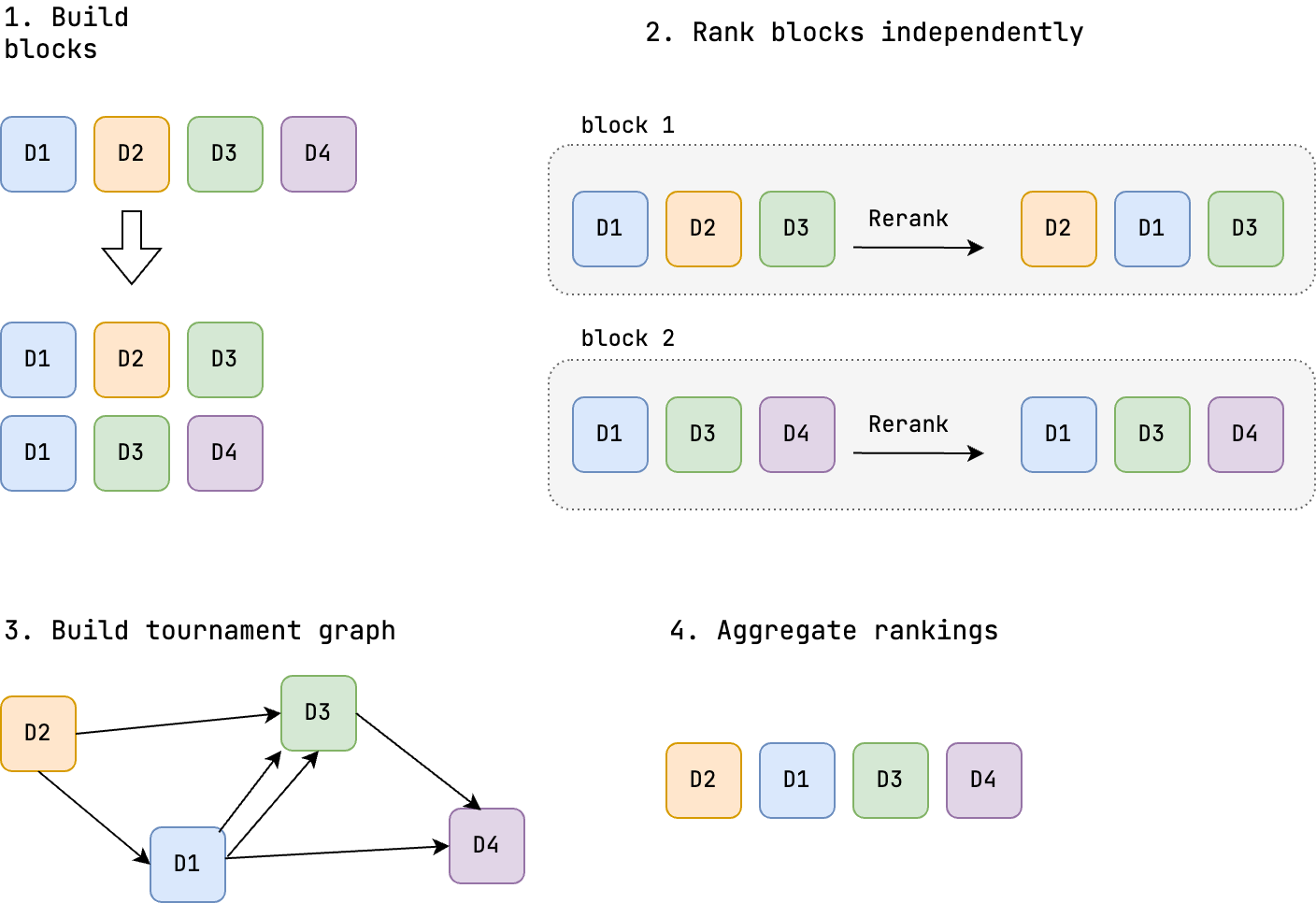}
    \caption{Joint Rank algorithm overview}
    \label{fig:joint}
\end{figure}

We propose a model-agnostic approach designed to efficiently construct batches ("blocks") of documents for parallel ranking in a single pass, without requiring subsequent calls to Listwise Reranker (e.g. LLM). Each block is carefully formed to overlap sufficiently with other blocks. Then, using a \textbf{listwise ranking model} for these overlapping blocks, we concurrently generate numerous implicit pairwise comparisons. The purposeful overlap between blocks ensures global connectivity, creating a \textit{tournament graph} from which a complete ranking can be derived effectively. The algorithm overview shown on \cref{fig:joint}.

Robust methods exist to reconstruct a complete ranking, including Elo Rating \cite{elo1978}, PageRank \cite{pagerank}, Winrate \cite{winrate}, Rank Centrality \cite{rank_centrality} and others. Additionally we considered Egienvalues \cite{eigen} and Bradley-Terry \cite{bradleyterry1952}.

The critical design choice for the algorithm is the strategy for building blocks, as it determines the overlap and ensures sufficient global connectivity in the tournament graph.

\subsection{Block Construction Designs}

We use the terminology and notation of experimental design here; the core principle is \textbf{blocking} - separating initial set into multiple subsets.
Other basic principles of experimental design are randomization (random elements in block as well as their random order within block), replication (each item occurs in several blocks), and balance (constant block size $k$, constant occurence $\lambda$ of each pair in blocks, constant $r$ replicas of each item) \cite{montgomery2017design}. Particular families of blockings are called block designs. In this work, we focus on incomplete block designs, meaning the blocks do not contain all elements.

\paragraph{Connected Block Design}

Two blocks are considered \textit{linked} if they share at least one common item. A block design is said to be \textit{connected} if, for any pair of items, there exists either a single block containing both items or a sequence of linked blocks such that the first block contains one item and the last block contains the other.

\paragraph{Covering Design}

To guarantee the reconstruction of a linear order on a set, every pair of items should co-occur in at least one block. Such a configuration is referred to as a \textit{covering design}.

\paragraph{Naive Sliding Window}

This approach simply takes adjacent overlapping windows in the sequence (e.g., with a sliding window of size $k$ moving forward by $<k$ steps). The main disadvantage of such approach is that the information flow is strictly limited to adjacent blocks. It implies that highly relevant documents found late in the ranking process cannot easily propagate relevance globally.
This design is connected and may be balanced with specific restrictions on parameters set and connecting first and last block.
This design relies on the initial order of elements and might be useful if we have good first-stage retriever. In this setup, overlapping can be managed more carefully by including the first and last elements of each window in the overlapped part.

\paragraph{Balanced Incomplete Block Designs}

In Balanced Incomplete Block Design \cite{montgomery2017design}, each pair of elements occurs in exactly $\lambda$ blocks. Each block contain distinct items. Each item presented in $r$ blocks, the block size $k$ is constant. PRP-AllPair \cite{llmpairwise} is a specific BIBD case ($k=2$). However, with increasing $k$ BIBD requires less blocks to consider.

BIBD is covering, balanced design. Redundancy is controlled via $\lambda$ and $r$.
Downsides of BIBD is that the total number of blocks is still large and it may be challenging to build and doesn't exist for certain parameters set.

\paragraph{Randomized Block Design}

Another alternative is randomly selecting subsets of documents to form blocks. Though simple to implement, random selection may yield insufficient comparability and connectivity, no balance guarantee, possibly leading to disjoint clusters.

\paragraph{Regular Equi-Replicate Block Design}

Regular Equi-Replicate Block Design (EBD) must satisfy following properties:

Consider $k$ is block size, $r$ number of replicates for each item. $v$ is total number of items, $b$ is block count.

\begin{itemize}
    \item Each block contain exactly $k$ items
    \item Every item replicated in exactly $r$ blocks
    \item All items in block are distinct
\end{itemize}

Consequently, 
\begin{equation}
 vr = bk \label{eq:vrbk}   
\end{equation}

\paragraph{Partially Balanced Incomplete Block Designs}

Partially Balanced Incomplete Block Designs \cite{bose1952classification} satisfy all the properties of Regular Equi-Replicate Block Designs but provide greater balance in terms of the connectivity among nodes in the corresponding comparison graph.

PBIBD($m$) defined by $m$ association classes, where every 2 items are $i$-th associates $i \in [1..m]$. Every $i$-th associates appears in exactly $\lambda_i$ blocks. Each item has exactly $n_i$ $i$-associates and the number $n_i$ does not depends on item. Another invariant is $p_{jk}^i$ -- for any pair of $i$-th associates, it represents the number of items common to the $j$-th associates of the one item and the $k$-th associates of the another.

Although PBIBDs provide more balanced designs, they have limitations regarding possible parameter choices, as PBIBDs only exist for specific combinations of ($v$, $r$, $b$, $k$). Moreover, no general algorithm works for all possible valid parameters sets, and finding a valid parameter set itself might be a challenge.

\subsection{Remarks on design for ranking problem}

In the context of the ranking problem, it is evident that the design must be connected to allow the estimation of at least a partial order; otherwise, the associated comparison graph will have more than one connected component. Connectivity alone may not be sufficient to ensure the recovery of a full linear order, but requires much less blocks compared to covering design. The guarantee of covering design sufficience is only valid if all pairwise comparisons are consistent, i.e., if the comparison relation is transitive and a-/anti-symmetric, conditions that may not always be satisfied by predicted comparisons. Using predictions, we should always treat resulting graph as tournament graph rather than comparison graph \cite{sortingpredictions}.

Although not formally proven here, we generally believe that, to minimize the ranking estimation error, the design should ideally be covering, balanced, randomized, and redundant.

However, given that obtaining rankings via ML systems is computationally expensive, we propose relaxing these requirements by recommending the use of a connected equi-replicate design or a connected PBIBD. In our research, we compared Randomized, EBD, PBIBD and Sliding Window designs.

\paragraph{Design implementations}
\label{ebdimp}
Implementing Randomized EBD is simple: concatenate $r$ independent shuffles of the initial sequence, then partition the result into blocks of $k$ elements. If $v$ is not divisible by $k$, a restriction must be added to ensure that blocks containing elements from adjacent sequences do not have repeated items. While EBD is not guaranteed to be connected, randomized construction makes it empirically highly probable (\cref{syn_connectivity}), and it is straightforward to sample another blocking if the initial one fails connectivity validation. Additionally, randomized construction makes it insensitive to initial order by design.

We also consider two algorithms for constructing PBIBD(2):

\begin{itemize}
    \item Latin-Square PBIBD \cite{bose1952classification}
    \item Triangular Association Scheme PBIBD \cite{bose1952classification}
\end{itemize}

A Latin Square can be constructed under the following conditions: $b = 2k$, $r = 2 \overset{\eqref{eq:vrbk}}{\implies} v = k^2$. It is created by arranging all items in a $k \times k$ matrix and considering both the rows and columns as blocks. In such design, each block become linked to exactly $k$ other blocks.

A Triangular PBIBD is constructed under the conditions: $v = b(b-1)/2$, $r = 2 \overset{\eqref{eq:vrbk}}{\implies} k = b-1$. The construction algorithm is detailed in \cite{bose1952classification}. Of particular interest in Triangular PBIBD is that any pair of blocks is linked.

\subsection{Efficiency}

The ranking of each block is independent and can be executed in parallel.
Block construction is highly efficient, with latency that can be considered negligible compared to the latency of an LLM call. For large $v$ or $b$, random sampling time may become noticeable; however, it can be pre-computed offline once. Similarly, rank aggregation algorithms, such as PageRank, run significantly faster than LLM calls. 
As a result, the overall latency is effectively dominated by the latency of a single reranker call.
When using an API, simultaneous parallel calls may be impacted by RPS throttling from the provider. While investigating such cases is beyond the scope of this study, we have documented real latency in seconds in \cref{tab:trec2019top100} (TogetherAI provider \footnote{\url{https://www.together.ai/}}) and \cref{tab:trec2019top1000} (OpenAI provider \footnote{\url{https://platform.openai.com/}}). However, in dedicated instances, RPS throttling will not influence latency as the number of parallel calls increases, provided the instance has sufficient capacity to handle all requests.

Complexity comparisons with other approaches are shown in \cref{tab:complexity}. Most entries were reused from \cite{tournament_rank}. For different methods, the meaning of hyperparameters is different. We did not include the complexity of TDPart, as the authors claimed it to be $O(N)$\cite{top_down}; however, this approximation does not account for the early stopping at $k$ or the window size $w$ parameters.

\begin{table}[h]
    \centering
    \caption{Complexity comparative analysis}
\begin{adjustbox}{width=\linewidth}
\begin{tabular}{lccc}
\toprule
\textbf{Methods} & \textbf{Time Complexity} & \textbf{No. Docs to LLM} & \textbf{Inferences} \\
\midrule
Pointwise & $O(1)$ & $N$ & $N$ \\
PRP-AllPair & $O(1)$ & $N^2 - N$ & $N(N-1)/2$ \\
Setwise.heapsort$(c,k)$ & $ O(k \log_{c}N)$ & $ ck \log_{c}N$ & $O(k \log_{c}N)$ \\
RankGPT-Sliding$(w,s)$ & $O(\frac{N}{s})$ & $N\frac{w}{w-s}$ & $\frac{N}{s}$ \\
TourRank$(K,r)$ & $O(K - 1)$ & $2r\cdot N$ & $\sum\limits_{k \in [1..K-1]}{rN_k}$ \\
JointRank$(k,r)$ & $O(1)$ & $r \cdot N$ & $rN/k$ \\
\bottomrule
\end{tabular}
    \label{tab:complexity}
\end{adjustbox}
\end{table}

\section{Synthetic Experiments}

Before evaluating on real-world data, we conducted extensive experiments on synthetic data to optimize hyperparameters and gain a comprehensive understanding of each algorithm's performance under various settings.

Our experiments suggest EBD and PBIBD blocks construction leads to superior comparison graph connectivity and robustness in global ranking inference. Among the tested approaches, PageRank consistently demonstrates the best performance in recovering accurate global rankings.

Increasing block size and blocks count increases quality of final ranking.

\subsection{Oracle reranker}
\paragraph{Experiment Setup}

We create a list of $v$ elements and assign exponential relevance values  ranging from $2^1$ to $2^v$, respectively. The list is then shuffled, and our goal is to recover the global order using an Oracle listwise reranker. The Oracle listwise reranker is assumed to output elements in their correct order for each block. We run this reranker on each block and subsequently attempt to reconstruct the global order using the pairwise comparisons generated from the block rankings.

For complete ranking recovery, we used tournament algorithms implemented by \cite{Ustalov:25} and open-source \footnote{\url{https://github.com/erensezener/rank-centrality}} implementation of Rank Centrality.

\paragraph{Evaluating block designs and rank aggregations}

First, we evaluate each block design and rank aggregation algorithm.
Since Latin-Square and Triangular PBIBDs can only be constructed under specific parameter restrictions (with each requiring different parameter combinations, making joint evaluation impossible), we consider two setups: \(v = 55\) for Triangular PBIBD and \(v = 100\) for Latin-Square PBIBD.
We sample input 1000 times for each combination of block design and rank aggregation algorithm and average results after.

\begin{table}[h]
    \centering
    \caption{Best result per design with $v = 55$, $k = 10$ and $b = 11$}
\begin{adjustbox}{width=\linewidth}
\begin{tabular}{llrrrr}
\toprule
\textbf{Design} & \textbf{Aggregation} & \textbf{Blocks Count} & \textbf{NDCG@10} & \textbf{Block Size} \\
\midrule
Triangular & PageRank & 11 & 0.87 & 10 \\
EquiReplicate & PageRank & 11 & 0.86 & 10 \\
SlidingWindow & Average winrate & 11 & 0.81 & 10 \\
Random & Average winrate & 11 & 0.74 & 10 \\
\bottomrule
\end{tabular}
    \label{tab:strategy_results}
\end{adjustbox}
\end{table}

\begin{table}[ht]
    \centering
    \caption{Aggregations results with Triangular PBIBD with $v = 55$ $k = 10$ and $b = 11$}
\begin{adjustbox}{width=\linewidth}
\begin{tabular}{llrrrr}
\toprule
\textbf{Aggregation} & \textbf{Design} & \textbf{Blocks Count} & \textbf{NDCG@10} &  \textbf{Block Size} \\
\midrule
PageRank & Triangular & 11 & 0.87 & 10 \\
Elo rating & Triangular & 11 & 0.85 & 10 \\
Average winrate & Triangular & 11 & 0.82 & 10 \\
Rank Centrality & Triangular & 11 & 0.77 & 10 \\
Eigen & Triangular & 11 & 0.11 & 10 \\
Bradley-Terry & Triangular & 11 & 0.10 & 10 \\
\bottomrule
\end{tabular}
    \label{tab:bibd_algorithm_results}
\end{adjustbox}
\end{table}

\begin{table}[ht]
    \centering
    \caption{Best result per design with $v=100$, $k=10$, $b=20$}
\begin{adjustbox}{width=\linewidth}
\begin{tabular}{llrrrrr}
\toprule
\textbf{Design} & \textbf{Aggregation} & \textbf{Blocks Count} & \textbf{NDCG@10} & \textbf{Block Size} \\
\midrule
Latin & PageRank & 20 & 0.76 & 10 \\
EquiReplicate & PageRank & 20 & 0.75 & 10 \\
SlidingWindow & PageRank & 20 & 0.68 & 10 \\
Random & PageRank & 20 & 0.62 & 10 \\
\bottomrule
\end{tabular}
    \label{tab:strategy_results2}
\end{adjustbox}
\end{table}

\begin{table}
    \centering
    \caption{Aggregation results for Latin $v=100$, $k=10$, $b=20$}
\begin{adjustbox}{width=\linewidth}
 \begin{tabular}{llrrr}
\toprule
\textbf{Aggregation} & \textbf{Design} & \textbf{Blocks Count} & \textbf{NDCG@10} & \textbf{Block Size} \\
\midrule
PageRank & Latin & 20 & 0.76 & 10 \\
Elo rating & Latin & 20 & 0.72 & 10 \\
Average winrate & Latin & 20 & 0.68 & 10 \\
Rank Centrality & Latin & 20 & 0.62 & 10 \\
Eigen & Latin & 20 & 0.06 & 10 \\
Bradley-Terry & Latin & 20 & 0.06 & 10 \\
\bottomrule
\end{tabular}
    \label{tab:bibd_algorithm_results2}
\end{adjustbox}
\end{table}

From \cref{tab:strategy_results}, we observe that the PBIBD constructed using the triangular association scheme achieves the best result, closely followed by the Equi-Replicate Block Design, which produces nearly similar scores. By contrast, the Naive Sliding Window and Random approaches demonstrate noticeably lower performance. \Cref{tab:bibd_algorithm_results} further indicates that the choice of rank aggregation method is critical: PageRank performs best when the design is balanced. However, for imbalanced blocks, the Average Winrate method outperforms others, as seen in \cref{tab:strategy_results} for the Sliding Window and Randomized Block Design.

Similarly, in the experiment with \( v = 100 \) using the Latin Square construction (\cref{tab:strategy_results2}, \cref{tab:bibd_algorithm_results2}), the PBIBD again yields the highest nDCG scores, with the Equi-Replicate design trailing by only one point.

Thus, both experiments consistently demonstrate the superior performance of PBIBDs. However, since it may be impossible or challenging to construct a PBIBD for certain parameter combinations due to structural constraints, we recommend using the Equi-Replicate Block Design in such circumstances, as it offers only a small reduction in performance.

In both experimental setups, PageRank consistently emerges as the best method for estimating complete rankings. Nonetheless, in certain cases, the simple average winrate can serve as a practical alternative: when paired with a suitable block design, this straightforward method achieves decent performance, trailing only slightly behind PageRank while being substantially easier to implement.

Eigenvalues and Bradley-Terry are highly sensitive to the chosen parameters, performing well only with strongly connected graphs and did not work in our case.

\paragraph{Effects of blocks count and different aggregators}

The aim of this experiment is to examine how the number of blocks affects the final ranking quality. We select the Equi-Replicate design with parameters \( v = 100 \) and vary the number of blocks \( b \) within the range \( [10, 100] \). Block size is fixed \(k=10\).

\begin{figure}
    \centering
    \includegraphics[width=\linewidth]{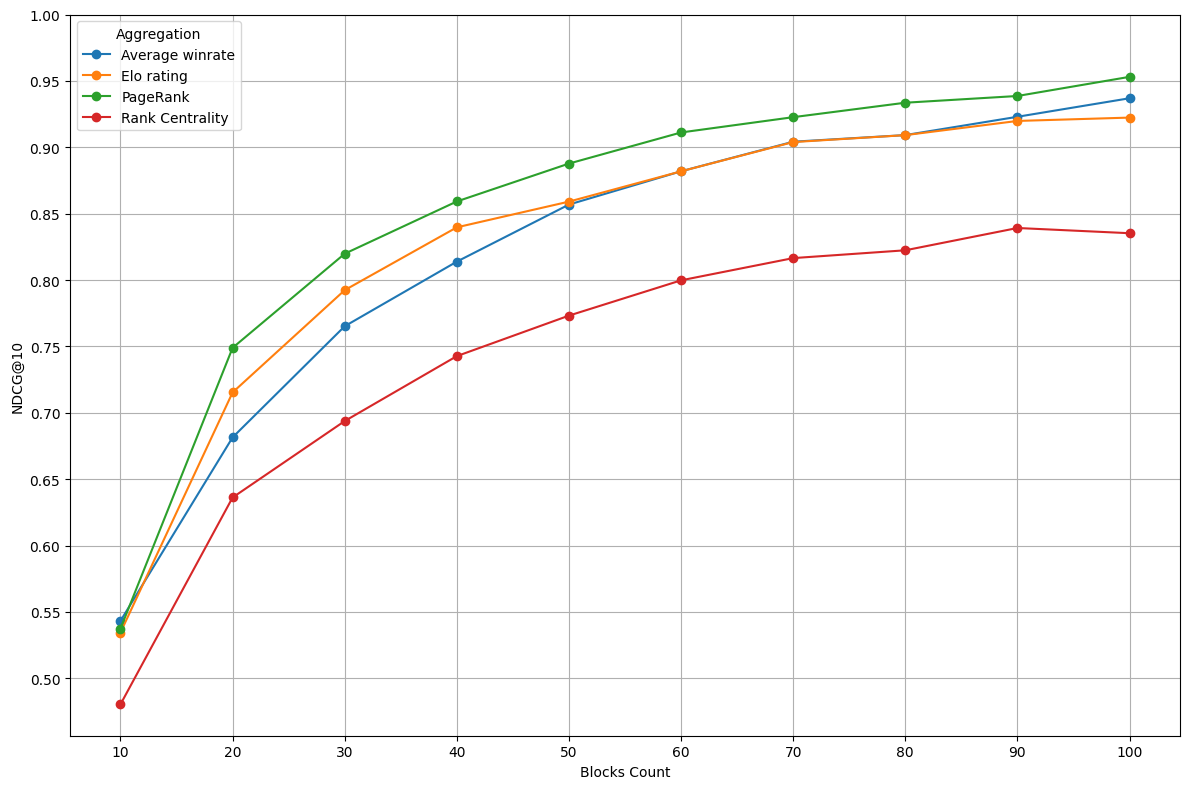}
    \caption{Blocks Count vs $nDCG@10$ for Aggregators (Equi-Replicate Design $v=100$)}
    \label{fig:estimators}
\end{figure}

Interestingly, the results presented in \cref{fig:estimators} again highlight that, when utilizing a well-constructed block design, even simple pairwise comparison ranking methods, such as winrate, achieve strong results -- only slightly behind those obtained by PageRank. This further emphasizes that the careful choice of block design can compensate for simplicity in the rank aggregation.

\paragraph{Evaluating different block size and block count configurations on larger candidate set}

\begin{figure}
    \centering
    \includegraphics[width=\linewidth]{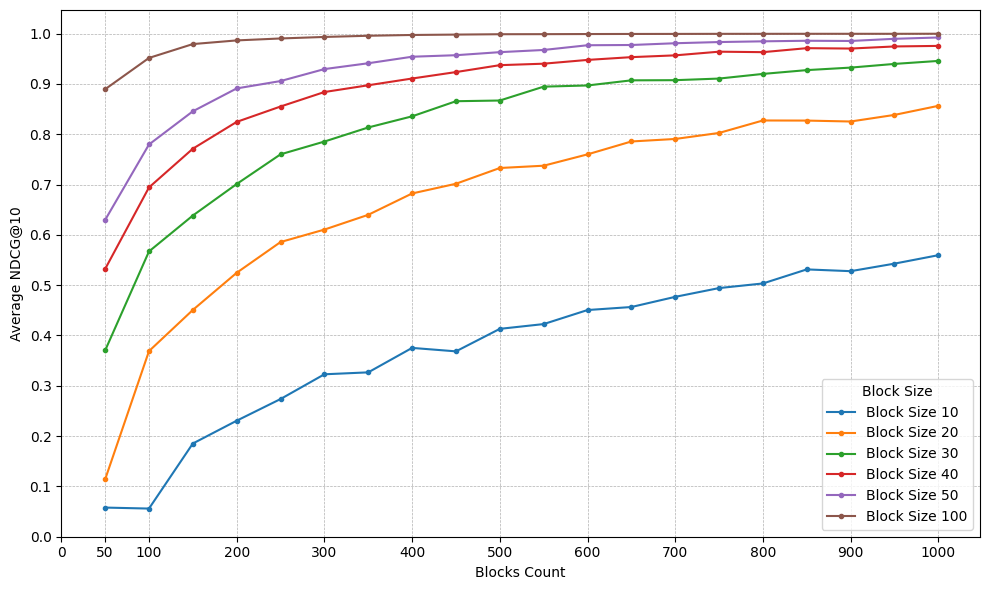}
    \caption{nDCG@10 for EBD-PageRank $v=1000$, $b\in[50..1000]$} 
    \label{fig:pagerank_ndcg_1000}
\end{figure}

\begin{figure}
    \centering
    \includegraphics[width=\linewidth]{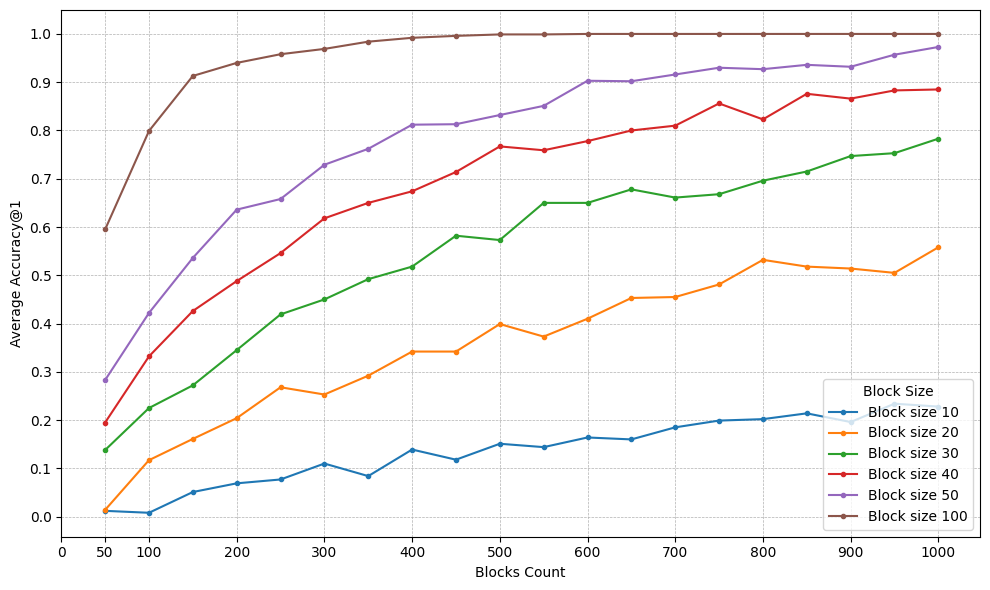}
    \caption{Accuracy@1 for EBD-PageRank $v=1000$, $b\in[50..1000]$} 
    \label{fig:pagerank_r_1000}
\end{figure}

We applied the Equi-Replicate \& PageRank approach and performed experiments to rank 1000 shuffled elements, testing various block sizes \((10,20,30,40,50,100)\) and block counts \([50..1000]\). Note that for certain parameter combinations, \eqref{eq:vrbk} has no integer solutions for $r$, leading to some items being underrepresented. In such cases we adhered to the implementation described in \cref{ebdimp}, but excluded the last blocks. Here, we focused on two performance metrics:

\begin{itemize}
    \item \( nDCG@10 \) (\cref{fig:pagerank_ndcg_1000})
    \item \( Accuracy@1 \) (the ability to correctly rank the most relevant item at position one) (\cref{fig:pagerank_r_1000})
\end{itemize}

Our experimental results reveal that both increasing the block size and increasing the number of blocks have a positive effect on ranking accuracy. However, block size has a significantly stronger impact on performance. This can be explained by the fact that the number of comparisons generated scales linearly with block count but scales quadratically with block size, since each block contributes \( k(k-1)/2 \) comparisons.

Consequently, to effectively rerank a longer sequence using smaller block sizes, we need a large number of blocks -- often exceeding the size of the candidate pool itself.

\subsection{Design Coverage}
\label{syn_connectivity}

To evaluate the design's ability to cover possible pairs, we consider the following statistics:

\begin{itemize}
    \item Rate of Direct Comparison Coverage: The percentage of pairs \((a, b)\) where both items \(a\) and \(b\) are presented together in at least one block.
    \item Rate of Second-Order Comparison Coverage: The percentage of pairs \((a, b)\) that are either directly included in a block or connected through an intermediary item \(c\). Specifically, \(c\) satisfies \((a, c)\) and \((c, b)\) within blocks, with \(a > c; c > b\) or \(a < c; c < b\).
    \item Average Node Degree: For each item \(a\), we calculate the average number of distinct items \(b\) such that \(a\) and \(b\) appear together in at least one block (average degree in the corresponding comparison graph).
    \item Minimum and Maximum Node Degrees: The lowest and highest node degrees in the comparison graph, indicating the least and most connected items.
    \item Co-Occurrence Rates: The average and maximum co-occurrence counts, measuring how frequently each pair of items appears together in the same block.
    \item Connectivity rate: The average rate of connectivity (1 if design is connected, 0 otherwise).
\end{itemize}

From \cref{tab:covering_stats100} and \cref{tab:covering_stats55}, we observe that PBIBD provides the best coverage for the same number of blocks (\(b\)) and block size (\(k\)). Additionally, this design is highly balanced in terms of item co-occurrence and node degrees. Such balance ensures fair representation of items and their combinations, which benefits subsequent rank aggregation methods, as demonstrated in \cref{tab:strategy_results} and \cref{tab:strategy_results2}. 

In contrast, Randomized Block Design shows the poorest statistics. However, even with less optimal designs such as Random and EBD, increasing \(b\) and \(k\) improves coverage. In conclusion, we recommend using PBIBD when feasible or EBD with larger \(b\) and \(k\), as EBD imposes fewer parameter restrictions.

\subsection{Offline design calculation}

One practical application of the synthetic evaluations described above lies in the offline selection of the blocking.

EBD is a superset of PBIBD; however, PBIBD yields superior results. This implies that it is possible to traverse various EBD instances to identify the most balanced design, ideally one that meets the PBIBD properties. This approach is useful in cases where the input parameters do not satisfy the requirements to build any PBIBD using known algorithms. In such scenarios, EBD instances can still be explored, optimizing either design coverage statistics or direct ranking metrics such as NDCG through evaluation on synthetic data with oracle reranker. Once the most suitable EBD instance is identified using synthetic data, it can be cached an applied to all real input in future. Algorithms like tabu search \cite{glover1998tabu} can support this optimization process, although such experiments were not conducted in this work.

\begin{table*}
    \caption{Coverage statistics for $v=100$; average over 1000 runs}
 \begin{adjustbox}{width=\textwidth}
 \begin{tabular}{lrrrrrrrr}
\toprule
\textbf{Design} & \textbf{1-comparisons} & \textbf{2-comparisons} & \textbf{avg. degree} & \textbf{min. degree} & \textbf{max. degree} & \textbf{co-oc. mean} & \textbf{co-oc. max} & \textbf{conn. rate} \\
\midrule
Random $(k=10, b=20)$ & .167 & .451 & 16.52 & 0 & 43.71 & .18 & 3.1 & 1.0 \\
EBD $(k=10, b=20)$ & .173 & .453 & 17.18 & 15.36 & 18.00 & .18 & 2 & 1.0 \\
Latin $(k=10, b=20)$ & \textbf{.182} & \textbf{.555} & \textbf{18.00} & 18.00 & 18.00 & .18 & 1 & 1.0 \\

\midrule
Random $(k=10, b=40)$ & .306 & .765 & 30.29 & 1.86 & 59.56 & .36 & 4.1 & 1.0 \\
EBD $(k=10, b=40)$ & \textbf{.317} & \textbf{.815} & \textbf{31.39} & 27.00 & 35.18 & .36 & 3.3 & 1.0 \\

\midrule
Random $(k=20, b=20)$ & .543 & .907 & 53.76 & 5.43 & 85.48 & .77 & 5.3 & 1.0 \\
EBD $(k=20, b=20)$ & \textbf{.574} & \textbf{.940} & \textbf{56.78} & 50.45 & 63.43 & .77 & 4 & 1.0 \\

\bottomrule
\end{tabular}
\end{adjustbox}
    \label{tab:covering_stats100}
\end{table*}

\begin{table*}
    \caption{Coverage statistics for $v=55$; average over 1000 runs}
 \begin{adjustbox}{width=\textwidth}
 \begin{tabular}{lrrrrrrrr}
\toprule
\textbf{Design} & \textbf{1-comparisons} & \textbf{2-comparisons} & \textbf{avg. degree} & \textbf{min. degree} & \textbf{max. degree} & \textbf{co-oc. mean} & \textbf{co-oc. max} & \textbf{conn. rate} \\
\midrule
Random $(k=10, b=11)$ & .287 & .560 & 15.52 & 0 & 34.40 & .33 & 3.3 & 1.0 \\
EBD $(k=10, b=11)$ & .308 & .642 & 16.64 & 14.86 & 18.00 & .33 & 2 & 1.0 \\
Triangular $(k=10, b=11)$ & \textbf{.333} & \textbf{.717} & \textbf{18.00} & 18.00 & 18.00 & .33 & 1 & 1.0 \\

\midrule
Random $(k=10, b=22)$ & .491 & .830 & 26.55 & 4.64 & 43.68 & .67 & 4.5 & 1.0 \\
EBD $(k=10, b=22)$ & \textbf{.521} & \textbf{.874} & \textbf{28.15} & 24.02 & 32.22 & .67 & 3.6 & 1.0 \\
\bottomrule
\end{tabular}
\end{adjustbox}
    \label{tab:covering_stats55}
\end{table*}

\section{Information Retrieval Evaluation}
\begin{table*}
    \centering
    \caption{TREC-2019 reranker evaluation for top-100 obtained by bm25}
\begin{adjustbox}{width=\textwidth}
 \begin{tabular}{llrrrrr}
\toprule
\textbf{Model} & \textbf{Algorithm} & \textbf{nDCG@10} & \textbf{\#Inferences} & \textbf{Pro. tokens} & \textbf{Gen. tokens} & \textbf{Latency}\\
\midrule
BM25 & - & 50.58 & - & - & - & - \\
Oracle & - & 89.22 & - & - & - & - \\
\midrule
Mistral-7Bv0.3 & \textbf{JointRank}$(r=4, k=20)$ & 58.43 & 20 & 53829 & 418 & \textbf{1s} \\
Mistral-7Bv0.3 & FullContextListwise & 51.03 & 1 & 11569 & 101 & \textbf{1s} \\
Mistral-7Bv0.3 & SlidingWindow$(w=20,s=10)$ & 60.24 & 9 & 24313 & 189 & 4s \\
Mistral-7Bv0.3 & Setwise.heapsort$(c=20, k=10)$ & 58.79 & 20 & 42306 & 215 & 7s \\
Mistral-7Bv0.3 & TDPart$(k=10, w=20)$ & 56.74 & 9 & 20372 & 601 & 3s \\
Mistral-7Bv0.3 & TourRank-2 & *58.69 & 26 & 44585 & 1410 & 4s \\
\midrule
Mistral-24B-2501 & \textbf{JointRank}$(r=4,k=20)$ & 69.10 & 20 & 50600 & 362 & \textbf{2s}\\
Mistral-24B-2501 & FullContextListwise & 61.76 & 1 & 10286 & 47 & \textbf{2s} \\
Mistral-24B-2501 & SlidingWindow$(w=20,s=10)$ & 65.86 & 9 & 22738 & 142 & 5s \\
Mistral-24B-2501 & Setwise.heapsort$(c=20, k=10)$ & 71.73 & 20 & 39910 & 51 & 8s \\
Mistral-24B-2501 & TDPart$(k=10, w=20)$ & 69.83 & 8 & 18277 & 147 & 4s \\
Mistral-24B-2501 & TourRank-2 & 65.03 & 26 & 40221 & 759 & 5s \\
\bottomrule
\end{tabular}
\end{adjustbox}
    \label{tab:trec2019top100}
\end{table*}

\begin{table*}
    \centering
    \caption{TREC-2019 reranker evaluation for shuffled top-1000 obtained by bm25}
\begin{adjustbox}{width=\textwidth}
 \begin{tabular}{llrrrrr}
\toprule
\textbf{Model} & \textbf{Algorithm} & \textbf{nDCG@10} & \textbf{\#Inferences} & \textbf{Pro. tokens} & \textbf{Gen. tokens} & \textbf{Latency} \\
\midrule
BM25 & - & 50.58 & - & - & - & - \\
Oracle & - & 96.40 & - & - & - & - \\

\midrule
gpt-4.1-mini & \textbf{JointRank}$(r=2, k=100)$ & 68.65 & 20 & 190k & 582 & \textbf{7s} \\
gpt-4.1-mini & \textbf{JointRank}$(r=3, k=100)$ & 70.88 & 30 & 284k & 887 & \textbf{8s} \\
gpt-4.1-mini & \textbf{JointRank}$(r=4, k=100)$ & 69.25 & 40 & 379k & 1134 & 10s \\

gpt-4.1-mini & FullContextListwise & 57.68 & 1 & 91k & 554 & 21s \\

gpt-4.1-mini & SlidingWindow$(w=100, s=50)$ & 73.50 & 19 & 180k & 828 & 62s \\
gpt-4.1-mini & SlidingWindow$(w=100, s=90)$ & 68.41 & 11 & 104k & 387 & 32s \\

gpt-4.1-mini & Setwise.heapsort$(c=100, k=10)$ & 71.39 & 28 & 246k & 110 & 64s \\

gpt-4.1-mini & TDPart$(k=10, w=100)$ & 71.40 & 13 & 107k & 385 & 14s \\
\bottomrule
\end{tabular}
\end{adjustbox}
    \label{tab:trec2019top1000}
\end{table*}
\subsection{TREC DL-2019 top-100 reranking}

For the initial evaluation, we use the TREC DL-2019 dataset \cite{trec2019}. We start by retrieving an initial top-100 list using BM25 and apply various reranking methods to this set. In these comparisons, we include several algorithms. \textbf{FullContextListwise} is single LLM call that includes the entire candidate set in the context. Other algorithms that process long sequences iteratively: \textbf{SlidingWindow} (RankGPT) \cite{rankgpt}, \textbf{Setwise.heapsort} \cite{setwise}, \textbf{TDPart} \cite{top_down}, \textbf{TourRank} \cite{tournament_rank}.
For fair comparisons, we align hyperparameters as follows: each LLM inference processes a maximum of 20 candidates at a time (commonly referred to as the window size). If the algorithm supports early stopping at top-$K$, we set \(K = 10\) as we are interested in \(nDCG@10\). For \textbf{JointRank} we used Regular Equi-Replicate Block Design and PageRank aggregation.

We also included non-functional metrics: \textit{\#Inferences} -- the average total number of LLM inferences required to process a single sample; \textit{Pro. tokens} -- the average total number of input tokens to the LLM; \textit{Gen. tokens} -- the average total number of tokens generated by the LLM; and \textit{Latency} -- the average time taken to process a single sample.

Finally, we include the results of an "Oracle reranker" -- a theoretically optimal reranker that leverages ground-truth relevance labels. Its score (89.22) represents the optimal achievable nDCG for this particular top 100 set, thus providing an upper bound reference point for our evaluation.

We took 2 models for the comparison: Mistral-7Bv0.3 \cite{mistral7b} and Mistral-24B-2501 \cite{mistral2501}.

Results in \cref{tab:trec2019top100} show that JointRank achieves the best latency, comparable to FullContextListwise, while significantly improving quality. Although some other algorithms demonstrate slight quality improvements over JointRank, their latency increases by at least twofold. \textit{*TourRank \cite{tournament_rank} reported a different score for Mistral-7B-Instruct-v0.2 (65.85). We use the score from our reproduction based on their source code, acknowledging responsibility for any potential issues in the setup.}

\subsection{TREC DL-2019 Long-context model performance on shuffled top-1000 reranking}

To better understand the impact of the initial first-stage ranking and to assess how effectively the FullContextListwise reranker retrieves relevant documents from large input windows, we conducted an additional experiment. Specifically, we take the initial top 1000 documents retrieved by BM25, randomly shuffle their order, and then pass this shuffled sequence to reranking algorithms. At this time, we allow to process 100 items at once for all algorithms except full-context.
For this evaluation, we used gpt-4.1-mini, as it demonstrated strong performance in the "search needle in a haystack" OpenAI-MRCR task \cite{gpt41}.

From \cref{tab:trec2019top1000}, we observe that the FullContextListwise method experiences a significant performance degradation when processing a large number of unordered passages. In addition, its latency increases sharply as the input size grows. The latency for JointRank also increased compared to single LLM call for the same input size -- well-likely due to throttling on the API side -- but it still maintains reasonable quality within an acceptable latency range. Setwise.heapsort, SlidingWindow and TDPart achieve slightly better quality but at the cost of increased latency. 
We selected $r = 3$ for JointRank as it offers a reasonable balance between quality and the number of parallel inferences. Increasing $r$ further did not lead to a quality improvement, which we interpret as the underlying model's capability to effectively rank 100 items at a time. Iterative approaches may achieve better quality, but process fewer items simultaneously in the final iterations.
We did not evaluate TourRank here, as the authors did not provide recommendations for scalability to arbitrary input size.

\subsection{BEIR evaluation}

Similar experiments were conducted on the BEIR \cite{thakur2021beir} benchmark.
Due to time constraints and limited model availability, comparisons are provided only for the gpt-4.1-mini model. Additionally, some algorithms listed in \cref{tab:trec2019top100} were excluded.
 
JointRank was evaluated in two configurations: processing a maximum of 10 documents at a time and a maximum of 20 documents at a time. This choice was made because BEIR containing subsets with significantly larger documents than TREC DL-2019. As observed in \cref{tab:beir_results}, reducing the block size improves both quality and latency for such input.
For other algorithms (TDPart and SlidingWindow), reducing the window size to 10 drastically increases latency, so the evaluation of these configurations was skipped.

JointRank with parameters $r=2, k=10$, uses the Latin-Square PBIBD and achieves the lowest latency, while TDPart offers better quality.

\begin{table*}
\centering
\caption{BEIR top-100 reranking obtained by bm25. nDCG@10 and latency in seconds per dataset.}
\begin{adjustbox}{width=\textwidth}
\begin{tabular}{l*{9}{cc}}
\toprule
\textbf{Algorithm} & 
\multicolumn{2}{c}{\textbf{trec-covid}} & 
\multicolumn{2}{c}{\textbf{robust04}} & 
\multicolumn{2}{c}{\textbf{webis-touche2020}} & 
\multicolumn{2}{c}{\textbf{scifact}} & 
\multicolumn{2}{c}{\textbf{signal1m}} & 
\multicolumn{2}{c}{\textbf{trec-news}} & 
\multicolumn{2}{c}{\textbf{dbpedia-entity}} & 
\multicolumn{2}{c}{\textbf{nfcorpus}} &
\multicolumn{2}{c}{\textbf{Avg.}}
\\
 & NDCG & Lat. & NDCG & Lat. & NDCG & Lat. & NDCG & Lat. & NDCG & Lat. & NDCG & Lat. & NDCG & Lat. & NDCG & Lat. & NDCG & Lat. \\
\midrule
BM25 & 59.47 & & 40.70 & & 44.22 & & 67.89 & & 33.04 & & 39.52 & & 31.80 & & 32.18 & & 43.60 & \\
Oracle 
& 97.48 & 
& 83.72 & 
& 91.41 & 
& 92.65 & 
& 62.00 &
& 82.66 & 
& 68.91 &
& 54.39 & 
& 79.15 &
 \\

\midrule
gpt-4.1-mini
& & & & & & & & & & & & & & & &  & & \\
\midrule

JointRank$(r=2,k=10)$
& 80.03 & 2.6
& 58.29 & 2.4
& 42.40 & 2.1
& 74.98 & 2.6 
& 30.47 & 2.0 
& 48.46 & 2.3
& 41.43 & 1.9 
& 35.16 & 2.3 
& 51.40 & \textbf{2.3}
\\

JointRank$(r=4,k=20)$
& 83.78 & 4.3
& 58.30 & 6.6
& 30.99 & 5.0
& 69.18 & 3.1 
& 31.43 & 2.7 
& 48.94 & 5.4 
& 40.75 & 2.8 
& 36.05 & 2.8 
& 49.93 & 4.1
\\

TDPart$(k=10,w=20)$ 
& 84.45 & 6.7 
& 60.13 & 5.9
& 38.02 & 4.9
& 78.84 & 3.8
& 33.99 & 4.4
& 50.51 & 5.5
& 44.77 & 3.7 
& 37.06 & 3.5
& \textbf{53.47} & 4.8
\\

SlidingWindow$(w=20,s=10)$ 
& 84.61 & 11.5
& 58.96 & 14.0 
& 37.94 & 12.5
& 78.28 & 10.7
& 32.75 & 8.9
& 49.34 & 13.5 
& 44.62 & 8.7 
& 36.67 & 7.6
& 52.90 & 10.9
\\

\bottomrule
\end{tabular}
\end{adjustbox}
\label{tab:beir_results}
\end{table*}

\section{Future Work}

Transforming block ranking into pairwise comparisons may result in a loss of information regarding the importance of each comparison. Items close together in the ranked sequence likely have smaller relevance differences, while items farther apart often differ more significantly in relevance. We experimented with using the distance between items, divided by block size, as weights for PageRank, but this had no impact. However, if implicit scores (e.g., those derived from a score-based listwise ranker) are available, their difference could be utilized as weights for improved performance.
Another potential research direction is combining JointRank with iterative or calibrating approaches to leverage the strengths of both methods. Furthermore, exploring different designs and rank aggregation methods could provide additional insights.

\section{Limitations}

Only a few models for zero-shot reranking were included, excluding supervised approaches. General recommendations, such as increasing balance, redundancy, and coverage to enhance final quality, are discussed, but formal proofs are not provided.

BEIR evaluation does not contain comparisions of all algorithms and different LLMs.

\section{Acknowledgements}

Egor Bogomolov from JetBrains Research for initial review and valuable feedback.

\section{Conclusion}

This paper addressed the critical challenge of applying high-quality listwise rerankers to large candidate sets within latency-sensitive information retrieval applications. We introduced JointRank, a model-agnostic framework that avoids the limits of model input size and the high latency associated with sequential processing inherent in many existing large-set reranking techniques. Our evaluations, which include both synthetic benchmarks, TREC DL-2019 and BEIR, confirmed the viability of this approach. Synthetic tests validated the importance of structured block designs and identified PageRank as a consistently effective aggregation strategy. On the TREC DL-19 benchmark, JointRank demonstrated a strong balance, achieving significant quality improvements over full-context reranking baseline (especially on large, unordered sets where long-context models faltered) while maintaining significantly lower latency compared to iterative methods. Hyperparameters for JointRank provide flexibility in balancing the trade-off between computational cost and ranking quality. In the end, JointRank offers a practical solution for deploying powerful but context-limited listwise rerankers in real-time systems, such as RAG or interactive search, closing the gap between advanced relevance modeling and the strict needs of low-latency environments.
\printbibliography

\end{document}